\documentclass{aastex631}

\usepackage{float}
\usepackage{amsmath}
\usepackage{enumitem}
\usepackage{upgreek}
\usepackage{longtable}

\submitjournal{PSJ}

\begin{document}

\title{Retrievals Applied To A Decision Tree Framework Can Characterize Earth-like Exoplanet Analogs}
\correspondingauthor{Amber Young}
\email{Amber\_Young86@nau.edu, ambervbritt@gmail.com}

\author[0000-0003-3099-1506]{Amber V. Young}
\affiliation{Northern Arizona University \\
527 S Beaver St. \\
Flagstaff, Arizona 86011, USA}
\affiliation{NASA Goddard Space Flight Center, Greenbelt, MD, 20771}

\author[0000-0003-2273-8324]{Jaime Crouse}
\affiliation{University of Maryland, College Park, MD, 20742}
\affiliation{NASA Goddard Space Flight Center, Greenbelt, MD, 20771}
\affiliation{NASA NExSS Virtual Planetary Laboratory, Seattle, WA, 98195, USA}


\author{Giada Arney}
\affiliation{NASA Goddard Space Flight Center, Greenbelt, MD, 20771}

\author{Shawn Domagal-Goldman}
\affiliation{NASA Goddard Space Flight Center, Greenbelt, MD, 20771}

\author{Tyler D. Robinson}
\affiliation{University of Arizona, Tucson, Arizona 85721, USA}

\author[0000-0003-2052-3442]{Sandra T. Bastelberger}
\affiliation{University of Maryland, College Park, MD, 20742}
\affiliation{NASA Goddard Space Flight Center, Greenbelt, MD, 20771}
\affiliation{Integrated Space Science and Technology Institute, Department of Physics, American University, Washington DC}




\begin{abstract}
Exoplanet characterization missions planned for the future will soon enable searches for life beyond our solar system. Critical to the search will be the development of life detection strategies that can search for biosignatures while maintaining observational efficiency. In this work, we adopted a newly developed biosignature decision tree strategy for remote characterization of Earth-like exoplanets. The decision tree offers a step-by-step roadmap for detecting exoplanet biosignatures and excluding false positives based on Earth's biosphere and its evolution over time. We followed the pathways for characterizing a modern Earth-like planet and an Archean Earth-like planet and evaluated the observational trades associated with coronagraph bandpass combinations of designs consistent with The Habitable Worlds Observatory (HWO) precursor studies. With retrieval analyses of each bandpass (or combination), we demonstrate the utility of the decision tree and evaluated the uncertainty on a suite of biosignature chemical species and habitability indicators (i.e., the gas abundances of H$_2$O, O$_2$, O$_3$, CH$_4$, and CO$_2$). Notably for modern Earth, less than an order of magnitude spread in the 1-$\sigma$ uncertainties were achieved for the abundances of H$_2$O and O$_2$, planetary surface pressure, and atmospheric temperature with three strategically placed bandpasses (two in the visible and one in the near-infrared). For the Archean, CH$_4$ and H$_2$O were detectable in the visible with a single bandpass.  
\end{abstract}

\keywords{Exoplanets, Biosignatures, Direct Imaging, Spectral Retrievals}


\section{Introduction} 
\label{sec:intro}
We as a science community are approaching a new horizon of scientific discovery for exoplanets. With the thousands of exoplanet detections that have been made to date, we are uniquely positioned to take the next steps toward exoplanet characterization and evaluating the potential for life to exist outside our solar system. The outcome of these efforts will depend on our ability to identify planets with habitable conditions and to remotely analyze them for signs of life. The recent Decadal Survey on Astronomy and Astrophysics 2020 recommended a space-based $\sim$6m telescope operating at near infrared/optical/ultra-violet wavelengths as a means to search for Earth-like planets orbiting nearby sun like stars and characterize their atmospheres with direct imaging and spectroscopy \citep{NAS_2021}. While this mission, dubbed The Habitable Worlds Observatory (HWO) by NASA, may not launch until the late 2030s or early-mid 2040s, now is the time to develop life detection strategies such that these science driven objectives can inform the instrumentation and architecture of future exoplanet characterization missions like HWO. This is especially important given that the search for life will likely drive multiple challenging telescope architecture and instrument decisions (e.g., coronagraph design). 

To conduct a search for life via spectroscopic observations, we can extract planetary contextual information from exoplanet spectra and search for biosignatures. This can generally be done through atmospheric retrievals, which are inference analyses that statistically determine atmospheric states able to reproduce the spectral data taking into account sources of uncertainty, for example, from observational noise \citep{Madhusudhan_2018}. Coronagraph bandpasses may be limited to $\sim$ 10--20\% \citep{LUVOIR_Study_2018}. Therefore, spectrally characterizing potentially Earth-like planets via reflected light coronagraphy with HWO may require a piece-wise approach in which we observe discrete portions of the planet's spectrum at a given time. The placement of these bandpasses and determining the observational requirements necessary to make biosignature detections is crucial for developing a robust search strategy and making informed instrument recommendations. Community endeavours like the Confidence of Life Detection (CoLD) Scale \citep{CoLD_Scale_2012} and the Biosignature Standards of Evidence workshop \citep{Biosig_Standards_Evidence_2022} have laid the groundwork in the context of generating community discourse on the topic and ideas for reporting a level of confidence associated with a potential life detection. Additionally, prior studies have explored the aspect of planetary and atmospheric inference studies as a function of signal-to-noise ratio (SNR) and resolution \citep[e.g.,][]{Madhusudhan_2009,Benneke_2012,Line_2013,Feng_2018,Barstow_2020}. However, there is work to be done to connect observational strategies for life detection to observational considerations for specific facilities like HWO. 

Here, we present a newly developed decision tree framework that organizes the workflow of observations and outlines bandpass choices to conduct a search for Earth-like atmospheric biosignatures based on Earth through time. We test the feasibility of this framework by calculating how well it can constrain key atmospheric species, despite only covering part of the planet's spectrum, for modern and Archean Earth as examples of types of exoplanets we may someday observe. Specifically, the decision tree considers bandpasses placed to observe key gases for the characterization of habitability and search for biosignatures: atmospheric water vapor (H$_2$O), molecular oxygen (O$_2$), ozone (O$_3$), methane (CH$_4$) and carbon dioxide (CO$_2$). H$_2$O is required by all life as we know it and remotely detecting it in an exoplanet atmosphere can help identify the habitable targets amongst a broader set of terrestrial planets \citep[e.g.,][]{Mottl_2007}. O$_2$ is a byproduct of the dominant metabolism on modern Earth, oxygenic photosynthesis, and a key biosignature that will be sought on exoplanets \citep[e.g.,][]{Meadows_2018}. O$_3$ can provide UV shielding from stellar activity that can be potentially harmful to organic life, but is also a photochemical byproduct of O$_2$ and can remain detectable at lower O$_2$ abundances \citep{Schwieterman_2018}. Finally, CH$_4$ is included because it is the main product of biogenic processes like methanogenesis and has few false positive mechanisms in the context of terrestrial planet atmospheres \citep[e.g.,][]{Wogan_2020}. Additionally, CH$_4$ detected alongside CO$_2$ could provide chemical context for a more oxidizing environment that would necessitate a substantial CH$_4$ flux in order to maintain its presence in the atmosphere \citep[e.g.,][]{Krissansen-Totton_2018,Arney_2018,Thompson_2022}. Atmospheric pressure is included in the decision tree to rule out a low pressure atmosphere false positive for O$_2$ that can leave interpretation of O$_2$ in absence of a detection of CH$_4$ ambiguous for modern Earth \citep{Wordsworth_2014}. 

The decision tree includes observational pathways tuned to detecting each of the aforementioned biosignature chemical species, and also has built in logic to account for potential false positive scenarios. We studied two pathways (modern Earth and Archean Earth) outlined in the decision tree as a first step towards testing the utility of this strategy for characterizing observationally distinct Earth-like biospheres. We chose modern Earth because it is the most well studied inhabited environment and a common point of reference in many exoplanet habitability studies. In contrast, the Archean was an era of Earth's geological history spanning roughly 4.0 to 2.5 billion years ago (Gyrs) that was representative of an anoxic, (but still inhabited) environment highly distinct from modern Earth. Observationally, Archean Earth would have looked substantially different from modern Earth and in the decision tree we reflect that with the observations that are prioritized for each planetary case. Examining both of these pathways allows us to conduct robust trade studies using simulated observations and retrievals to guide instrument development in the context of future exoplanet direct imaging and characterization missions.

\section{Methods} 
\label{sec:meth}
We performed a series of atmospheric studies to test a proposed observational strategy to characterize potentially Earth-like exoplanets with the HWO. The main goal of this study was to assess the general feasibility of the decision tree framework, and highlight key trade-offs that could be used to inform the development of HWO. The proposed decision tree framework outlines a step-by-step workflow of observations, which are prioritized based on searching for Earth-like habitability markers, atmospheric biosignature species, and ruling out potential false-positives. Using the retrieval model, we executed each level of the decision tree strategy and made key atmospheric inferences. 

\subsection{The Decision Tree Observational Strategy}
\begin{figure}[H]
    \centering
    \includegraphics[width=\linewidth]{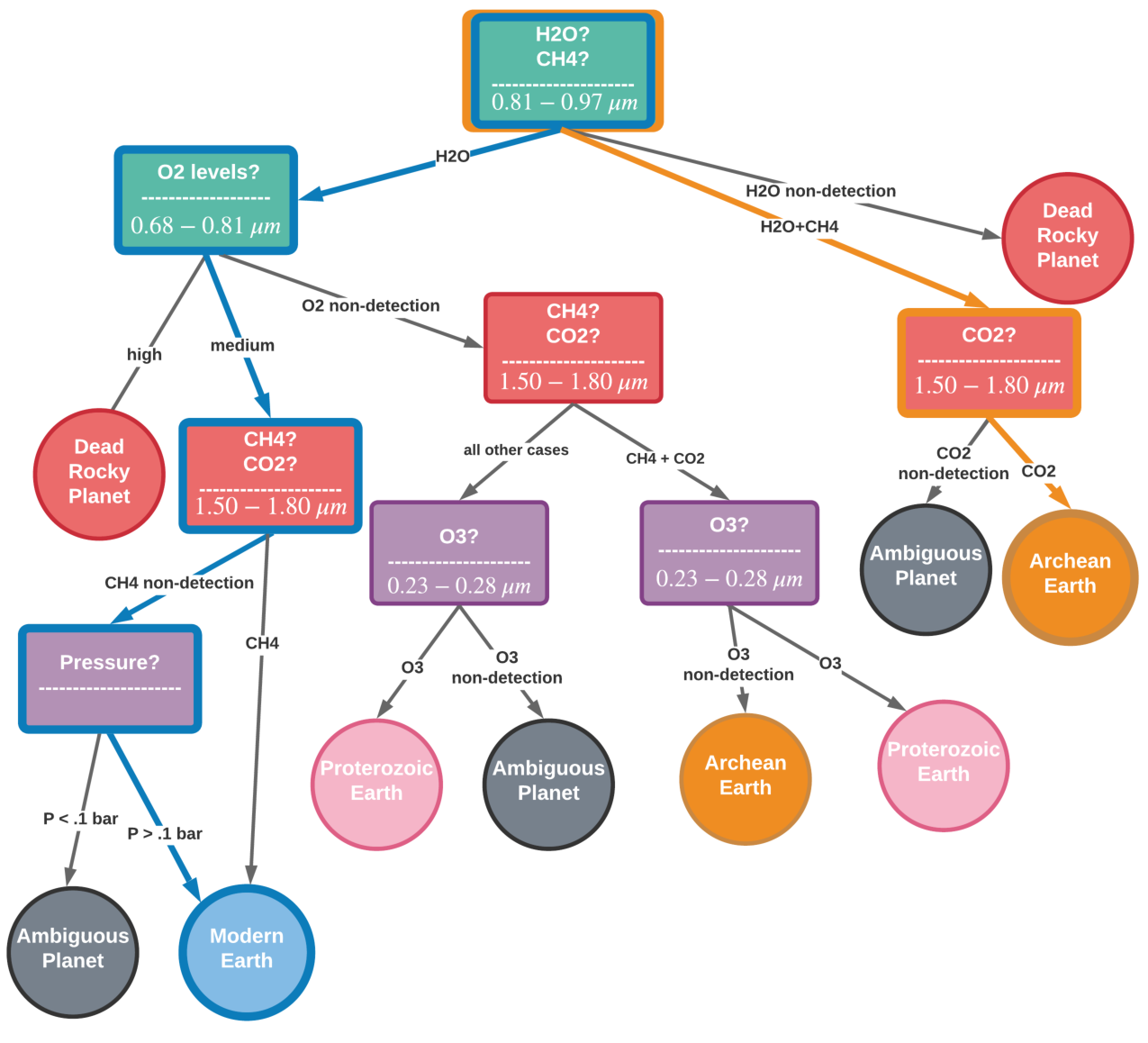}
    \caption{Cartoon schematic of the decision tree framework. The modern Earth and Archean Earth paths that are highlighted 
    in blue and orange respectively, were selected for exploration in our retrieval simulations as a first look at how this framework could work in observations of exoplanets. In general, the purple, green, and red boxes represent wavelength regions in the UV, visible, and NIR channels respectively. In practice, observations following the framework would be performed from the top to the bottom where at the base, observational targets could be categorized into the respective planetary scenarios. Ambiguous planetary scenarios are in dark grey, dead rocky cases are in red, Modern Earth-like cases are in blue, Proterozoic Earth-like cases are in pink, and Archean Earth-like cases are in orange.}
    \label{DT_diagram}
\end{figure}

This observational strategy offers a top level guideline for detecting exoplanet biosignature species and excluding false positives based on our current best understanding of Earth's biosphere and its evolution throughout geologic history. Outlined below (Figure \ref{DT_diagram}) is a cartoon diagram of our adopted decision tree showing the step-by-step process for characterizing Earth-like biosignatures. While this decision tree is tailored for the search for biosignatures on exoplanets analogous to Earth history, future iterations of HWO decision trees could be broadened to include other types of planets, including non-habitable ones. 
\newpage

In practice, the order of observations for a given high priority target would be executed from top to bottom where the color coding for each layer of the decision tree represents the following coronagraph channels split by wavelength ranges: visible from 0.4 -- 1.0\,$\upmu$m (green), near infrared from 1.50 -- 1.80\,$\upmu$m (red), and ultra-violet from 0.2--0.4\,$\upmu$m (purple). In practice, those observations that utilize different coronagraph channels could be performed in parallel. Hypothetically, the search would start with H$_2$O to establish the possibility of habitability. In that same observation that searches for H$_2$O, CH$_4$ could also be observed if it were present at abundances higher than modern Earth and more akin to Archean Earth with specifics to be determined by future work. If water or both species were to be detected, then one could move on to the next set of observations outlined in the decision tree. For example, if water vapor was detected but methane was not, one could perform a search for molecular oxygen with the 0.68--0.81\,$\upmu$m bandpass, as this would suggest a planet deficient in CH$_4$ and perhaps similar to modern Earth. If O$_2$ were detected, one could next search for CH$_4$ in the NIR to help rule out biosignature false positive scenarios for O$_2$ and search for an additional line of evidence for life. If CH$_4$ were not detected, obtaining constraints on atmospheric pressure are a useful way to rule out a low pressure atmosphere false positive scenario for O$_2$. Alternatively, if methane was detected along with the initial water vapor observation, a search for CO$_2$ could be conducted in the 1.50--1.80\,$\upmu$m bandpass to provide geophysical context for interpreting the CH$_4$ as a biosignature (or not), and this bandpass additionally has a CH$_4$ absorption feature that would improve the CH$_4$ constraint. Each observation could be performed sequentially until an end branch is reached in the decision tree. Those end branches then indicate planets of multiple types which could range from dead and rocky, to observationally ambiguous, to a potential biosignature detection characteristic of an era of Earth's history.

In this work, we focused on simulating habitable Earth-like planets around a sun-like star and use the decision tree to characterize both a modern Earth-sun twin and an Archean Earth-sun twin. However, the decision tree could be adapted to consider potential Earth-like planets around other host stars as well. For example, a modern Earth-like planet around an M dwarf or late K dwarf host star may produce more substantial amounts of CH$_4$ given the different photochemistries driven by the different UV spectra of these stars \citep[e.g.,][]{Segura_2005,Arney_2019}. With detectable levels of CH$_4$, O$_2$, H$_2$O, and an inferred global surface pressure above 0.1 bar, one could still classify that planet as modern Earth-like, given knowledge of these stars' photochemistries, but the optimal order of observations in a decision tree framework might differ. We focus on the G-type scenario here as a starting point, but developing iterations of the tree for other stars will be an important avenue for future work. 

\subsection{The Retrieval Model}
The open-source \texttt{rfast} model \citep{Robinson_2023_rfast_PSJ} was adopted for the atmospheric retrievals. It incorporates a radiative transfer forward model, an instrument noise model, and a retrieval tool to facilitate ``fast" investigations of exoplanet atmospheric remote sensing scenarios. The radiative transfer forward model can simulate both 1-D and 3-D views of an exoplanet in (1) reflected light, (2) emission, and (3) transit, and the atmospheric chemical and thermal state (including profiles of cloud properties) are the forward model inputs. The incorporated noise model is based on \citet{Robinson_2016}, and provides wavelength-dependent noise estimates for a variety of observing scenarios. Finally, the retrieval package employs \texttt{emcee}, a Bayesian sampling package \citep{Foreman_Mackey_2013}, which calls the aforementioned radiative transfer model while mapping out the posterior distribution for the atmospheric and planetary parameters used to fit a noisy observation.

In applications of the atmospheric retrieval model presented below, flat priors are used to bound parameter space and the likelihood function is given by,
\begin{equation}
  \log \mathcal{L} \propto -\frac{1}{2} \chi^{2} \ ,
\end{equation}
where $\chi^{2}$ is computed in the standard fashion given a set of simulated datapoints, associated Gaussian uncertainties, and a model prediction. The retrievals assume an isothermal temperature profile (representative of a atmospheric column-averaged temperature) with constant profiles assumed for the species mixing ratios. The decision tree framework is not formally Bayesian and our atmospheric retrieval approach simply maximizes the likelihood function within the bounds of the imposed flat priors. 

\subsection{Modern Earth \& Archean Earth Atmospheric Modeling}
Self-consistently calculated atmospheres were generated using the photochemical model incorporated as part of the \texttt{Atmos} tool \citep{arney_pale_2016,Arney_pale_2017,Teal_2022}. \texttt{Atmos} is a coupled 1-D photochemical-climate model that uses planetary inputs (e.g., chemical species mixing ratios, chemical species fluxes, gravity, and stellar spectrum) and associated chemical reactions to calculate the steady-state atmospheric profiles of gases, hazes (if applicable), pressure, temperature etc. In both the modern Earth and Archean Earth atmospheric scenarios, the default solar spectrum was used to model Earth-like cases relevant to reflected light observations \citep{Thullier_etal_2004}. The simulated modern Earth atmospheric case assumed a total fixed surface pressure of 1\,bar. The atmospheric state we simulated for the Archean Earth-like scenario represents a haze free case and was modeled with a CO$_2$ mixing ratio of 4\% and a CH$_4$/CO$_2$ ratio of 0.075, which is well below 0.1 where hazes are expected to form \citep{Trainer_2006}. Additionally, for the Archean Earth scenario, the solar spectrum was scaled to accurately reflect its wavelength-dependent insolation from 2.7 Gyrs ago \citep{Claire_2012}. 

\subsection{Simulated \texttt{rfast} Observations}
Reflected light observations in this study were modeled after the decadal direct imaging exoplanet mission concepts that will likely inform the future development of HWO. For the purpose of our study, we assume HWO's proposed exoplanet coronagraphic instrumentation will be based off the properties of the coronagraphs studied in the Large-UV-Optical-Infrared (LUVOIR) and Habitable Exoplanet (HabEx) observatory concept studies \citep{LUVOIR_Study_2018,HabEx_Study_2018}. This coronagraph instrument would include three separate coronagraph channels, labeled as ``ultra-violet" (with a wavelength range of 0.2--0.4\,$\upmu$m), ``visible" (0.4--1.0\,$\upmu$m), and ``near-infrared" (1.0--1.8\,$\upmu$m). Over this range, the observed spectrum is split into smaller spectral bandpasses. It is possible to conduct simultaneous observations in two of the different observational channels (i.e., bandpasses in the visible and NIR channels could be observed in parallel, while two bandpasses in a single channel must be observed in series). In this study, each observational channel is modeled at resolving powers  of 7, 140, and 70 for the UV, visible, and NIR respectively based on the LUVOIR design \citep{LUVOIR_Study_2018}. We then assume 20\% spectral bandpasses across each wavelength channel. Following the Modern Earth decision tree pathway (See Figure \ref{DT_diagram}) we have three main observational bandpasses of interest. (1) The 0.89\,$\upmu$m bandpass at 0.81--0.97\,$\upmu$m, which is capable of measuring H$_2$O abundances at Earth-like levels, as well as CH$_4$ abundances higher than modern levels. (2) The 0.75\,$\upmu$m bandpass at 0.68--0.81\,$\upmu$m, which was chosen to retrieve on atmospheric abundances of O$_2$ by encapsulating the strongest O$_2$ absorption feature in this range (the O$_2$ A-band) at 0.762\,$\upmu$m. (3) The 1.65\,$\upmu$m bandpass at 1.50--1.80\,$\upmu$m, which can be used to constrain (but not necessarily measure) CH$_4$ and CO$_2$ abundances. We simulated observations using each singular bandpass and subsequently modeled observations of bandpass combinations all at a nominal SNR of 10. Table  \ref{tab:ME_1sig} and Table \ref{tab:AE_1sig} of the Appendix include all the single and bandpass combinations that we simulated for characterizing modern Earth-like and Archean Earth-like atmospheric scenarios, respectively. Each observation consisted of a simulated noisy spectral dataset which was generated with randomized error bars and constant noise set at the shortest wavelength within a given bandpass combination. We adopted this formalism for modeling the observational noise from results outlined in the studies from the LUVOIR and HabEx decadal survey reports \citep{LUVOIR_Study_2018,HabEx_Study_2018}.

\newpage
\section{Results} 
\label{sec:res}
Retrieval results are analyzed for information inferred from spectral ``pieces" of modern Earth-like (Section \ref{ME_res}) and Archean Earth-like  (Section \ref{AE_res}) spectra comprised of 20\% spectral bandpasses according to their respective pathways on the decision tree. The modern Earth pathway (Figure \ref{DT_diagram} blue path) included the 0.75\,$\upmu$m, 0.89\,$\upmu$m, and 1.65\,$\upmu$m bandpasses and the Archean Earth pathway (Figure \ref{DT_diagram} orange path) included the 0.89\,$\upmu$m and 1.65\,$\upmu$m bandpasses. The following subsections highlight constraints for species volume mixing ratio abundances for a number of gases related to habitability or a biosignature search: O$_2$, H$_2$O, CO$_2$, O$_3$, and CH$_4$. It also includes the retrieved values for various atmospheric state parameters, including surface albedo, planetary mass/radius, surface pressure, and atmospheric temperature.

\subsection{The Modern Earth Pathway}
\label{ME_res}
\begin{figure}[H]
    \centering \includegraphics[scale=0.70]{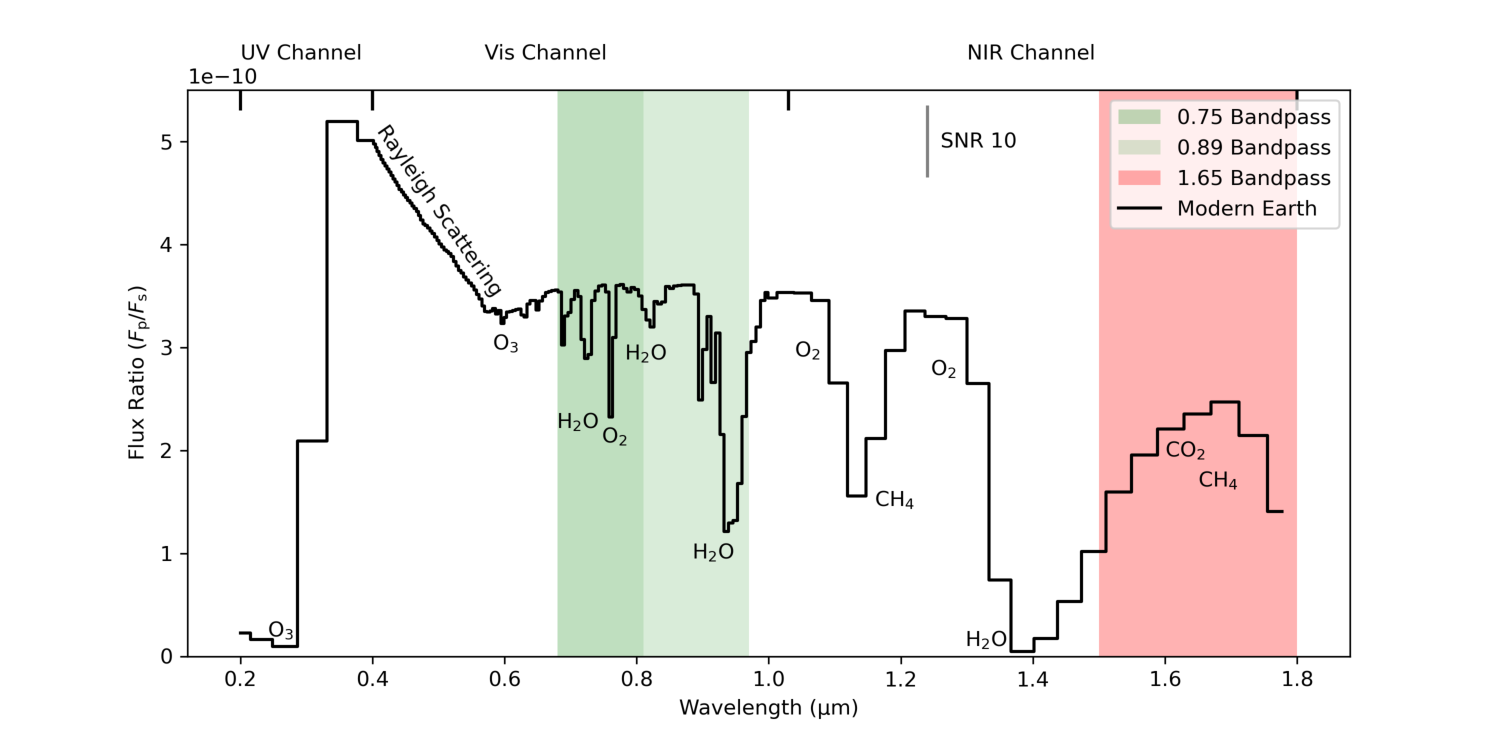}
    \caption{Simulated Reflected Light Spectrum of Modern Earth. Shown here is the reflected light spectrum of modern Earth with coronagraph wavelength channels indicated along the top for the UV (0.2--0.4\,$\upmu$m), the Vis (0.4--1.0\,$\upmu$m), and the NIR (1.0--1.8\,$\upmu$m). Individual 20\% bandpasses are indicated according to the decision tree strategy with the 0.75\,$\upmu$m bandpass (0.68--0.81\,$\upmu$m) dark green shaded region, the 0.89\,$\upmu$m bandpass (0.81--0.97\,$\upmu$m) green shaded region, and 1.65\,$\upmu$m bandpass (1.50--1.80\,$\upmu$m) red shaded region. The Rayleigh scattering slope and absorption features of key species are labeled in text along with an error bar scaling for an SNR of 10 relative to 0.68\,$\upmu$m. Note that atmospheric retrievals were performed on individual or a combination of these three bandpasses.}
    \label{Spectrum}
\end{figure}

Figure \ref{Spectrum} depicts the full spectrum of modern Earth which was retrieved on in pieces. Each spectral piece corresponds to a 20\% bandpass indicated in the decision tree and labeled by the midpoint of each spectral swath. In the 0.75\,$\upmu$m region, the O$_2$ A-band feature is present at 0.76\,$\upmu$m, and an H$_2$O feature is present at 0.72,\,$\upmu$m. The 0.89\,$\upmu$m bandpass includes two prominent H$_2$O features at 0.82\,$\upmu$m and 0.94\,$\upmu$m. Finally the 1.65\,$\upmu$m region contains weak CH$_4$ (at 1.65\,$\upmu$m) and CO$_2$ (at 1.6\,$\upmu$m) absorption features. The goal of this study was to determine the amount of planetary information that could be inferred from a given observation spanning one or a combination of these smaller spectral regions.

\begin{figure}[H]
    \centering
    \includegraphics[scale=0.70]{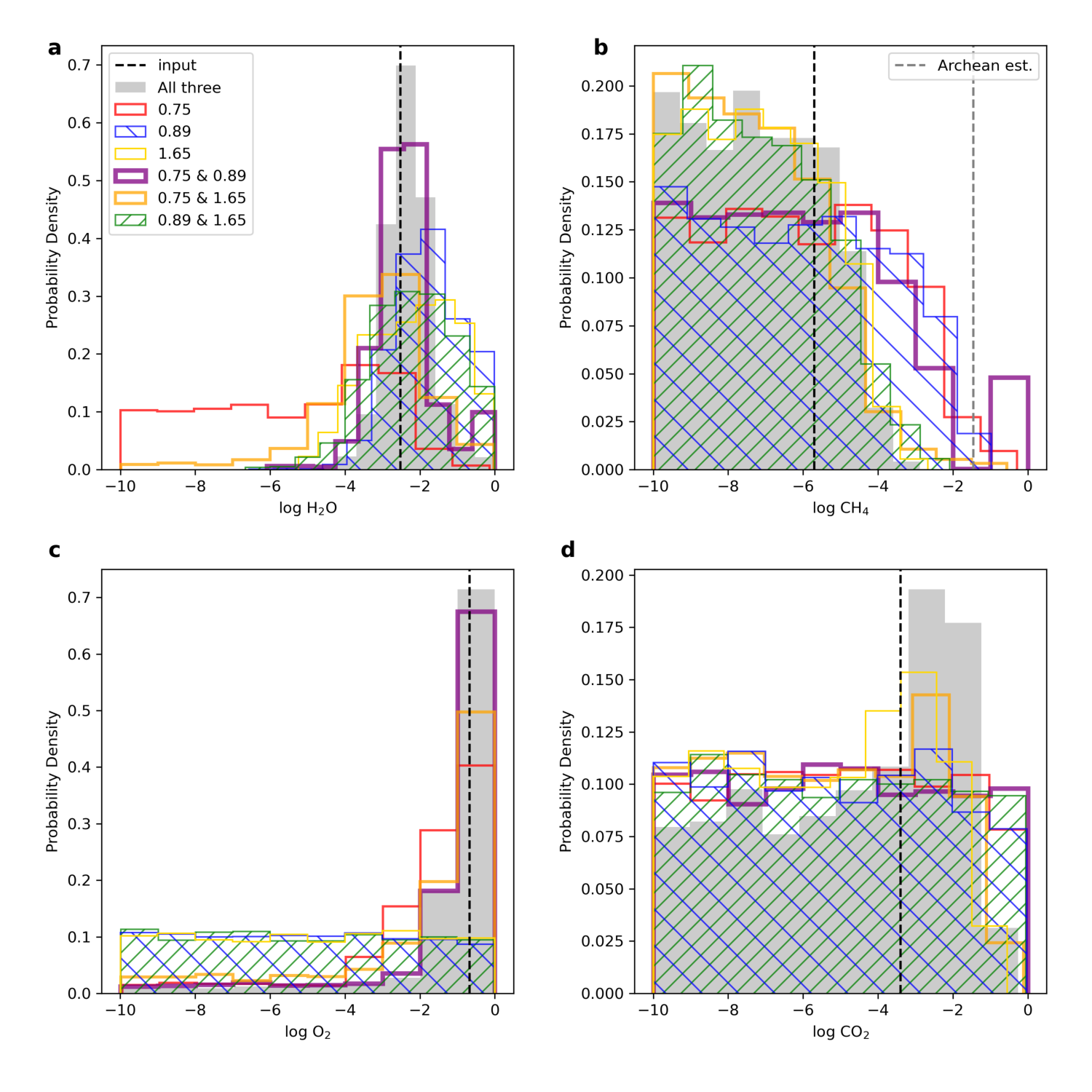}
    \caption{Posterior Distributions for Key Decision Tree Biosignature Species (modern Earth). \textbf{a}, Posterior distributions for the H$_2$O abundance derived from simulated SNR 10 observations at various combinations of wavelength coverage. The ``All three" combination encompasses the 0.89\,$\upmu$m bandpass, the 0.75\,$\upmu$m bandpass, and the 1.65\,$\upmu$m bandpass (grey filled). The singular observational bandpasses include the 0.75\,$\upmu$m bandpass (red), the 0.89\,$\upmu$m bandpass (blue hatched), and the 1.65\,$\upmu$m bandpass (yellow). The paired observations include the 0.75\,$\upmu$m \& 0.89\,$\upmu$m bandpass pair (purple), the 0.75\,$\upmu$m \& 1.65\,$\upmu$m bandpass pair (orange), and the 0.89\,$\upmu$m \& 1.65\,$\upmu$m bandpass pair (green hatched). For CH$_4$, O$_2$, and CO$_2$, black vertical dashed lines indicate the surface values given by the input atmospheric model. The input value for H$_2$O (also indicated by a vertical dashed line) is taken to be the column average mixing ratio. \textbf{b}, The same as \textbf{a} but for the CH$_4$ abundance. The grey vertical dashed line represents a modeled estimate for an Archean Earth-like abundance of CH$_4$ ($3.5 \times 10^{-2}$) from \citet{Robinson_Reinhard_2020} for comparison to the modern value. \textbf{c}, same as \textbf{a} and \textbf{b} but for the O$_2$ abundance. \textbf{d}, same as \textbf{a}, \textbf{b}, and \textbf{c}, but for the CO$_2$ abundance.} 
    \label{Biosig_spec}
\end{figure}

The breadth of information that can be inferred from spectral information is strongly dependent on the singular bandpass choice (or bandpass combination). Figure \ref{Biosig_spec} shows the retrieved marginal posterior distributions of gas mixing ratios for H$_2$O (Figure \ref{Biosig_spec}a), CH$_4$ (Figure \ref{Biosig_spec}b), O$_2$ (Figure \ref{Biosig_spec}c), and CO$_2$ (Figure \ref{Biosig_spec}d). 

\newpage
\noindent The retrievals show that observations gathering spectral information from all three bandpasses (labeled as ``All three") provide the best constraints for each of these gases, and the majority of atmospheric parameters that were retrieved on (this is described in more detail below). Notably, the 0.75\,$\upmu$m \& 0.89\,$\upmu$m bandpass combination provides excellent constraints on both the O$_2$ and H$_2$O abundance and is second only to the ``All three" combination in terms of the quality of the constraints. For CH$_4$ and CO$_2$, it is extremely difficult to constrain these species with any of the bandpass combinations. At best, we obtain upper limit constraints on the abundances of each of these species. Most of the posteriors for CH$_4$ and CO$_2$ exhibit distributions with a significant drop off at a given value not dictated by the statistical prior of that parameter making these upper limit constraints. In Figure \ref{Biosig_spec}b, a vertical grey dashed line representing an Archean Earth-like estimate of $3.5 \times 10^{-2}$ for the CH$_4$ abundance (adopted from \cite{Robinson_Reinhard_2020}) is included and demonstrates that we can only rule out high CH$_4$ scenarios with the atmospheric retrievals.

\begin{figure}[H]
    \centering \includegraphics[scale=0.74]{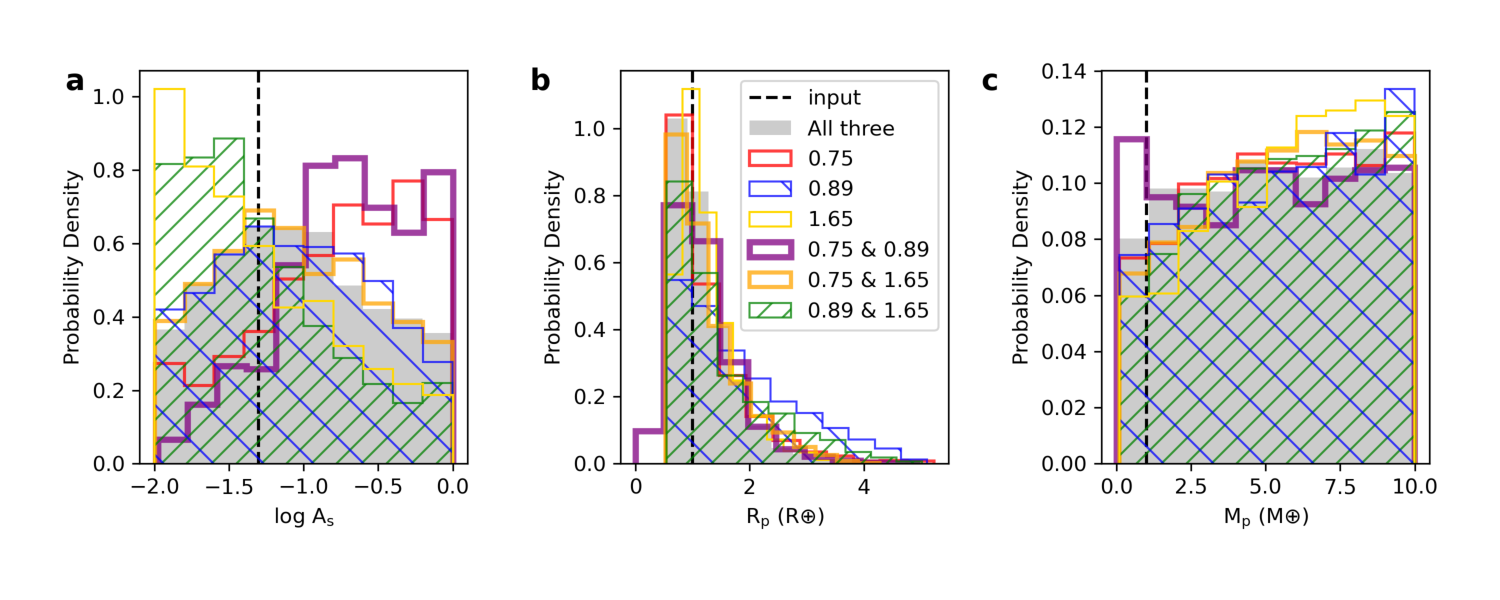}
    \caption{Posterior Distributions for Planetary Surface Albedo, Planetary Radius, and Planetary Mass. \textbf{a}, Posterior distributions for the log of the planetary surface albedo (A$_{\rm s}$) derived from simulated SNR 10 observations and all of the decision tree combinations of wavelength coverage. The input values for each parameter are indicated by a black vertical dashed line. \textbf{b}, Same as \textbf{a}, but for the planetary radius (R$_{\rm p}$). \textbf{c}, Same as \textbf{a} and \textbf{b}, but for the planetary mass (M$_{\rm p}$).}
    \label{physical_params}
\end{figure}

Figure \ref{physical_params} outlines the marginal posteriors for the log of the surface albedo (A$_{\rm s}$), the planetary radius (R$_{\rm p}$), and planetary mass (M$_{\rm p}$). We found that constraints on the surface albedo (Figure \ref{physical_params}a) are strongly dependent on the bandpass (or bandpass combination) choice. For planetary radius (Figure \ref{physical_params}b), we find the posteriors for all the observational cases exhibit distributions that are peaked closely to ``truth" (i.e., the original input value) and have statistical tails extending toward higher values. Planetary mass (Figure \ref{physical_params}c) goes unconstrained for all observational cases and is bound solely by the mass prior.

\begin{figure}[H]
    \centering \includegraphics[scale=0.7]{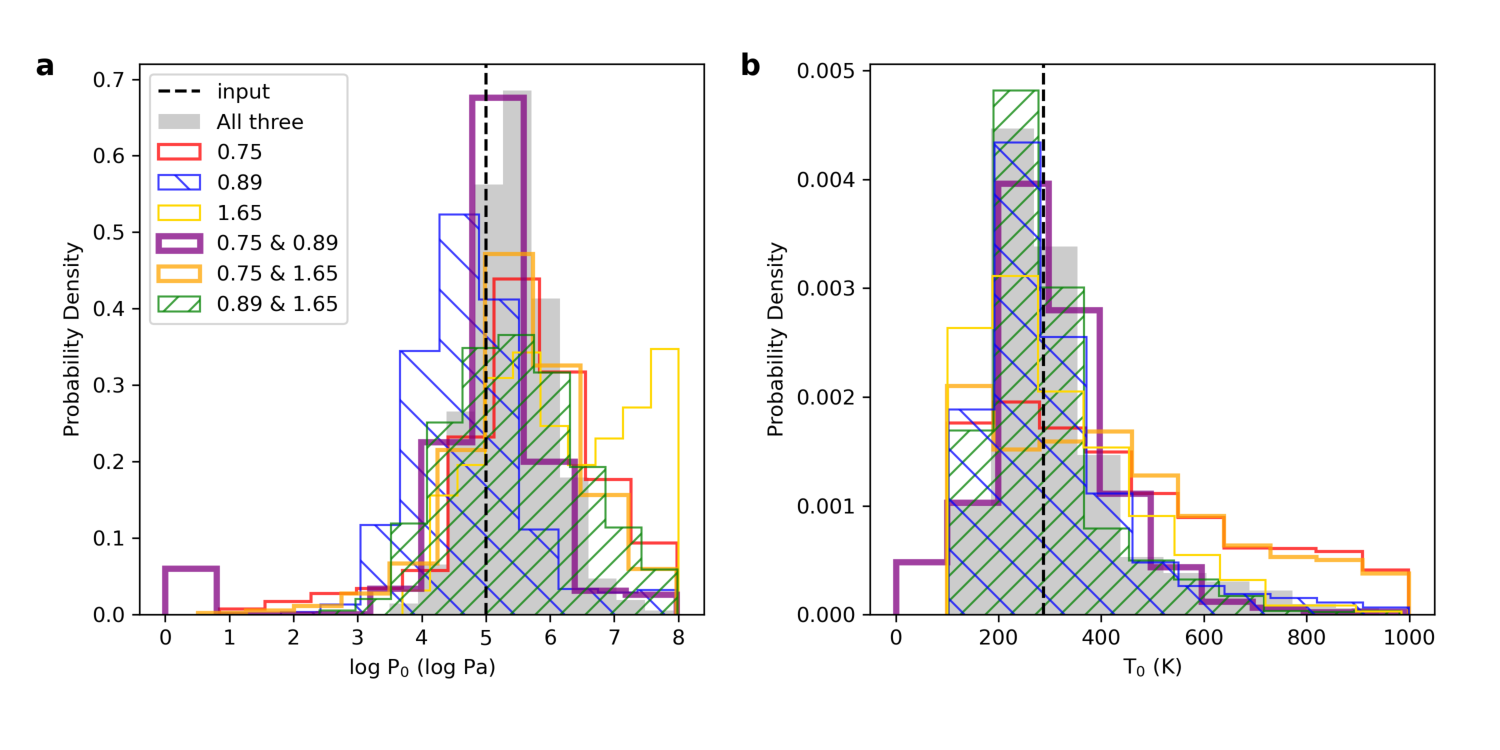}
    \caption{Posterior Distributions for Global Surface Pressure and Global Temperature. \textbf{a}, Posterior distribution for global surface pressure of the planet derived from simulated SNR 10 observations and all of the decision tree combinations of wavelength coverage. For each parameter, black vertical dashed lines indicate the surface values given by the input atmospheric model. \textbf{b}, Same as \textbf{a}, but for temperature. Note that the retrieved spread would be representative of temperatures at the surface and throughout the troposphere.}
    \label{thermo_params}
\end{figure}

Constraints on the surface pressure (P$_0$) and the characteristic atmospheric temperature (T$_0$) are shown in Figure \ref{thermo_params}. For surface pressure (Figure \ref{thermo_params}a), a majority of the distributions (most notably the ``All three" combination and the 0.75\,$\upmu$m \& 0.89\,$\upmu$m combination) have peaked distributions near truth and exhibit somewhat Gaussian behavior with tails on either side of the peak. In these retrieval analyses, the bandpasses were not selected to detect a Rayleigh slope for an Earth-like planet. However, pressure constraints can come from the widths of gas absorption bands due to broadening. 

\noindent In Figure \ref{thermo_params}b we also find that the constraints on atmospheric temperature are peaked near the surface temperature (288 K) and have extended tails toward higher temperatures for all the observational scenarios. The temperature constraints ultimately stem from temperature dependent opacities (mainly H$_2$O).

\begin{figure}[H]
    \centering
    \includegraphics[scale=0.60]{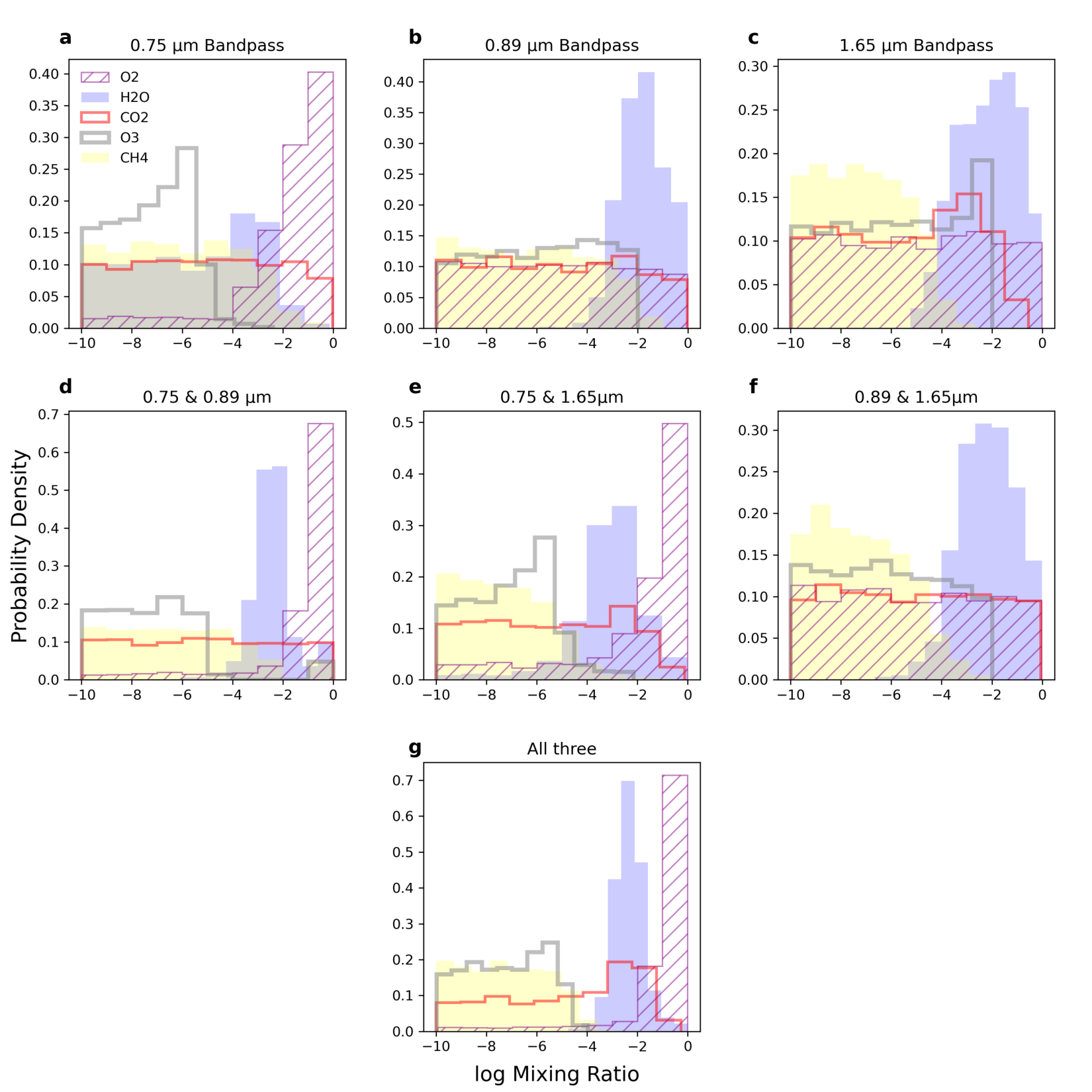}
    \caption{Biosignature Abundance Constraints Broken Down by Observational Bandpass Combinations. \textbf{a}, Chemical species abundance constraints for the 0.75\,$\upmu$m bandpass. The abundance constraints for O$_2$ (purple hatched), H$_2$O (purple filled), CO$_2$ (red), O$_3$ (grey), and CH$_4$ (yellow filled) are all shown. \textbf{b}, Abundance constraints for the same species but for the singular 0.89\,$\upmu$m bandpaas observation. \textbf{c} Constraints for the same species but for the singular 1.65\,$\upmu$m bandpass observation. \textbf{d}, Constraints for the same species but for the 0.75\,$\upmu$m \& 0.89\,$\upmu$m bandpass pair. \textbf{e}, Constraints for the same species but for the 0.75\,$\upmu$m \& 1.65\,$\upmu$m bandpass pair. \textbf{f}, Constraints for the same species but for the 0.89\,$\upmu$m \& 1.65\,$\upmu$m bandpass pair. \textbf{g}, Constraints for the same species but for ``All three" bandpasses combined (i.e., 0.75\,$\upmu$m \& 0.89\,$\upmu$m \& 1.65\,$\upmu$m bandpasses). }
    \label{bandpass_abund}
\end{figure}

Figure \ref{bandpass_abund} takes Figure \ref{Biosig_spec} and re-formats it such that the posterior distributions for all the atmospheric biosignatures are shown for an individual bandpass or combination of bandpasses in each subplot. This is to better highlight the bandpass combinations that may be optimal for detecting (or constraining) one or more biosignature species. For instance, the 0.89\,$\upmu$m bandpass (Figure \ref{bandpass_abund}b) provides a clear detection of H$_2$O (filled purple distribution) but provides no constraint on the O$_2$ abundance (purple hatched distribution). In order to get constraints on both H$_2$O and O$_2$ the  0.75\,$\upmu$m \& 1.65\,$\upmu$m (Figure \ref{bandpass_abund}e), the 0.75\,$\upmu$m \& 0.89\,$\upmu$m (Figure \ref{bandpass_abund}d), or the ``All three" (Figure \ref{bandpass_abund}g) bandpass combinations would be required.

\newpage

\newpage

\subsection{The Archean Earth Pathway}
\label{AE_res}
\begin{figure}[H]
    \centering
    \includegraphics[scale=0.7]{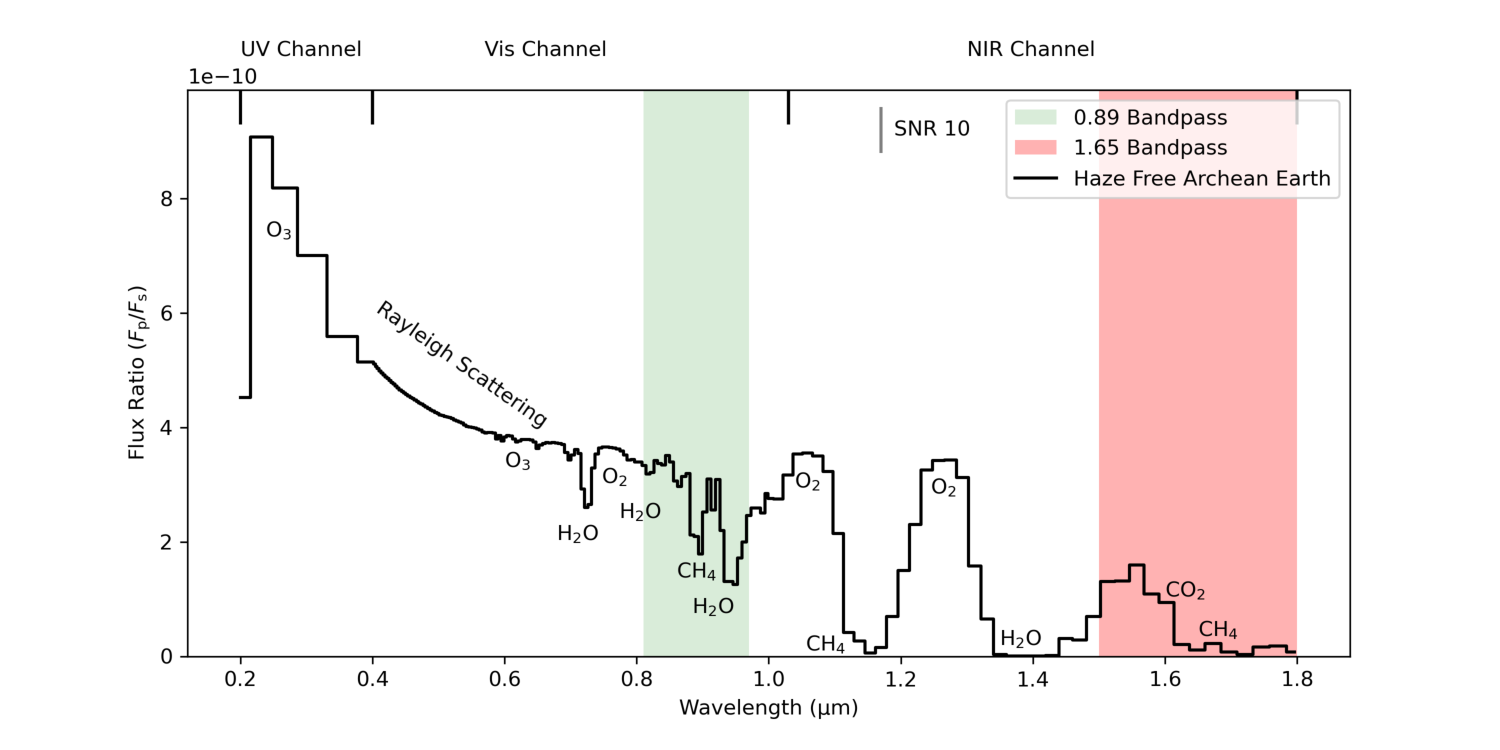}
    \caption{Modeled Haze free Archean Earth Spectrum. Shown here is the planet-to-star flux ratio vs wavelength and in black is the simulated spectrum for a haze free Archean Earth-like planet around the sun. The spectral bandpasses from the Archean Earth pathway on the decision tree are shown where the 0.89\,$\upmu$m bandpass is shaded in green and the 1.65\,$\upmu$m bandpass is shaded in red. The Rayleigh scattering slope and species absorption features are labeled in text. Also shown is an SNR 10 error bar scaling indicative of the simulated observations for each singular bandpass and combination.}
    \label{AE_Spectrum}
\end{figure}

For the Archean Earth pathway on the decision tree, there are two main bandpasses of interest, the 0.89\,$\upmu$m bandpass and the 1.65\,$\upmu$m bandpass. This has fewer observations in the path, because in this case the ``search for water" can occur in the same bandpass as the search for the primary biosignature, CH$_4$. This feature is detectable at CH$_4$ concentrations higher than those present on modern Earth, and that may occur on biospheres similar to Archean Earth whose primary producers generated CH$_4$ in an oxygen-poor atmosphere. This feature might also be detectable in an oxygen-rich atmosphere around M and K dwarfs \citep{Segura_2005, Arney_2019}, whose photochemisty allows for higher CH$_4$ abundance in the presence of oxygen; this will be treated in the decision tree framework in future studies. Figure \ref{AE_Spectrum} shows a noise-free Archean Earth spectrum generated from the aforementioned Atmos atmospheric profiles. The atmosphere used in our analysis does not include a haze, which may have been intermittently present during the Archean \citep[e.g.,][]{arney_pale_2016}. The 0.89\,$\upmu$m and 1.65\,$\upmu$m bandpasses are color coded in the green and red shaded regions respectively and the proposed wavelength regions for each coronagraph channel are marked along the top axis. Absorption features of key species are labeled along with the spectral slope due to Rayleigh scattering. Within each of the two bandpasses there are CH$_4$ features at 0.87\,$\upmu$m and 1.27\,$\upmu$m, which both provide key CH$_4$ abundance information in the atmospheric retrievals. Additionally, there is a CO$_2$ absorption feature encapsulated in the 1.65\,$\upmu$m bandpass at 1.6\,$\upmu$m and H$_2$O absorption features in the 0.89\,$\upmu$m bandpass at 0.82\,$\upmu$m and 0.94\,$\upmu$m. 

\begin{figure}[H]
    \centering
    \includegraphics[scale=0.6]{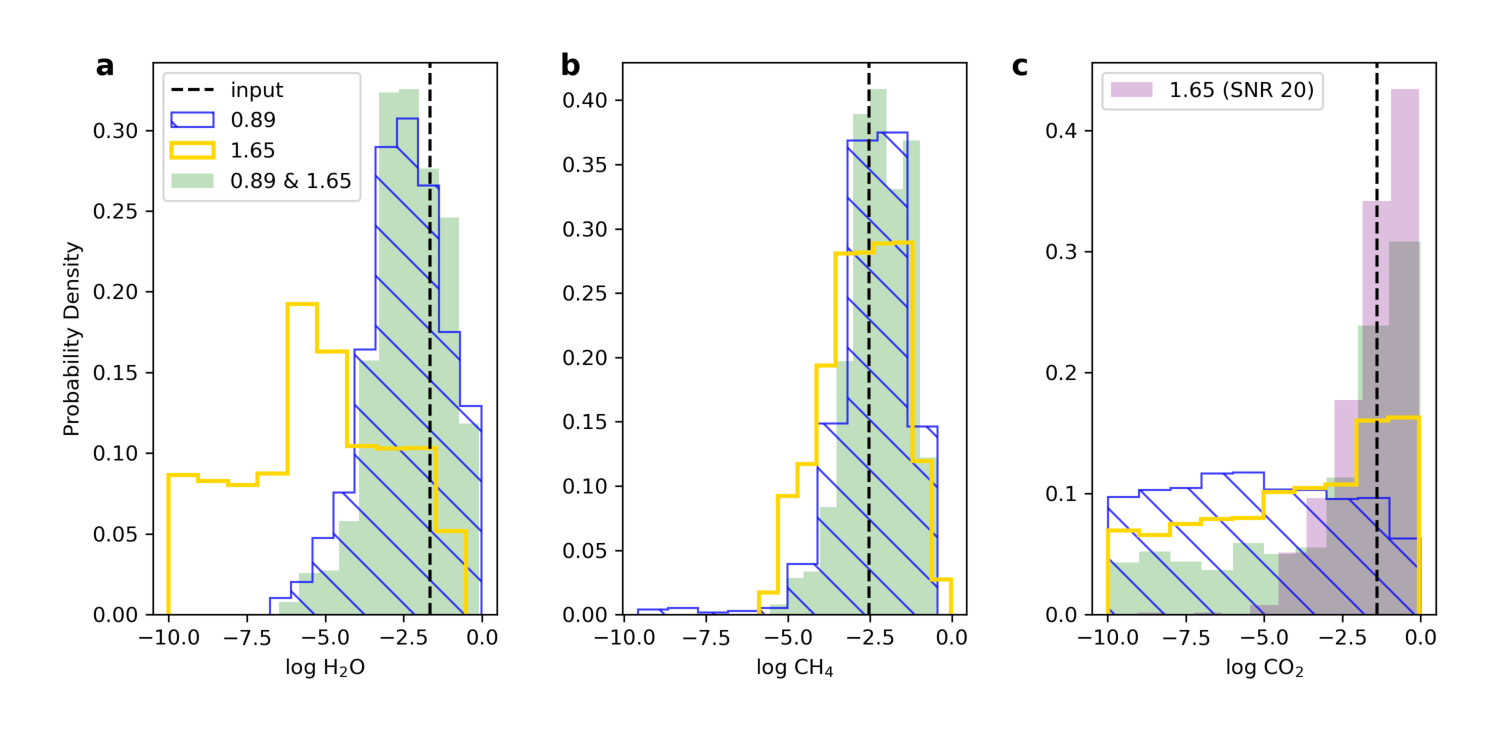}
    \caption{Posterior Distributions for Key Decision Tree Species (Archean Earth Pathway). \textbf{a}, Marginal posterior distributions for the H$_2$O abundance derived from the simulated SNR 10 observations covering two singular bandpasses, the 0.89\,$\upmu$m bandpass, the 1.65\,$\upmu$m bandpass, and a combination of the two. The posterior for the 0.89\,$\upmu$m bandpass is illustrated in the blue hatch distribution, the 1.65\,$\upmu$m posterior is the yellow distribution and the combination of the two is the green filled distribution. For CH$_4$ and CO$_2$, black vertical dashed lines indicate the surface values given by the input atmospheric model. The input value for H$_2$O (also indicated by a vertical dashed line) is taken to be the column average mixing ratio. \textbf{b}, The same as \textbf{a} but for the CH$_4$ abundance. \textbf{c}, The same as \textbf{a}, and \textbf{b}, but for the CO$_2$ abundance. An additional retrieved posterior is included for a simulated SNR 20 observation with the 1.65\,$\upmu$m singular bandpass (purple filled).}
    \label{AE_Biosig}
\end{figure}

Figure \ref{AE_Biosig} shows the resulting constraints on the H$_2$O, CH$_4$, and CO$_2$ abundances for the singular and combined bandpass observations. For CH$_4$ especially (Figure \ref{AE_Biosig}b), the overall constraints are much improved in comparison to the modern Earth case because of its increased abundance for this Archean Earth-like scenario. We also find here that the prominent CH$_4$ feature at 0.87\,$\upmu$m allows for the detection of CH$_4$ at the shorter wavelength bandpass compared to modern Earth. Because the CO$_2$ constraints are weaker compared to the H$_2$O and CH$_4$ constraints with the nominal SNR 10 observation, we simulated an additional SNR 20 observation with the 1.65\,$\upmu$m singular bandpass. We find (Figure \ref{AE_Biosig}c) at higher SNR the retrievals produced a more sharply peaked posterior distribution for CO$_2$ (Figure \ref{AE_Biosig}c purple filled curve). H$_2$O (Figure \ref{AE_Biosig}a) can be detected with the 0.89\,$\upmu$m \& 1.65\,$\upmu$m bandpass combination or the singular 0.89\,$\upmu$m bandpass.

\begin{figure}[H]
    \centering
    \includegraphics[scale=0.65]{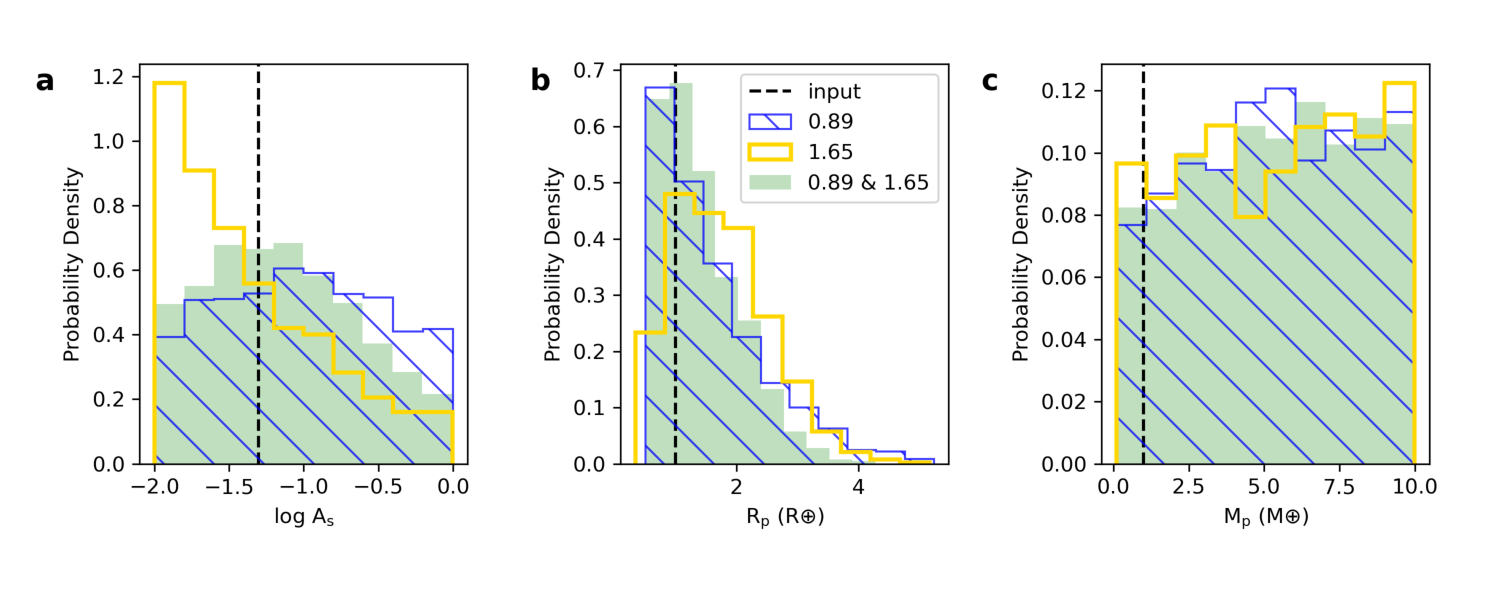}
    \caption{Posterior Distributions for Planetary Surface Albedo, Planetary Radius, and Planetary Mass (Archean Earth Pathway). \textbf{a}, Posterior distributions for the log of the planetary surface albedo (A$_{\rm s}$) derived from simulated SNR 10 observations and all of the decision tree combinations of wavelength coverage for the Archean Earth path. The input values for each parameter are indicated by a black vertical dashed line. \textbf{b}, Same as \textbf{a}, but for the planetary radius (R$_{\rm p}$). \textbf{c}, Same as \textbf{a} and \textbf{b}, but for the planetary mass (M$_{\rm p}$).}
    \label{AE_Physical}
\end{figure}

\begin{figure}[H]
    \centering
    \includegraphics[scale=0.65]{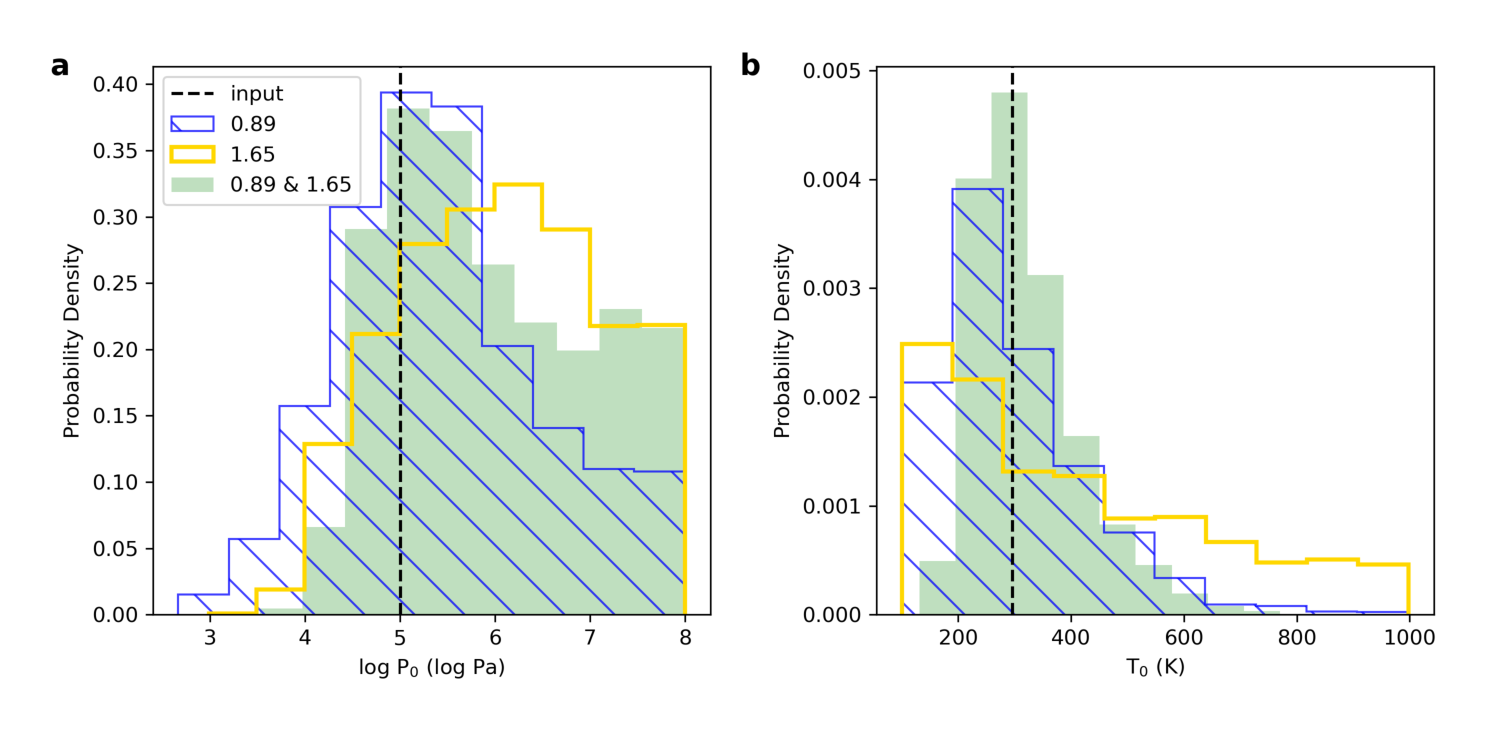}
    \caption{Posterior Distributions for Global Surface Pressure and Global Temperature. \textbf{a}, Posterior distribution for global surface pressure of the planet derived from simulated SNR 10 observations and all of the decision tree combinations of wavelength coverage for the Archean Earth path. For each parameter, black vertical dashed lines indicate the surface values given by the input atmospheric model. \textbf{b}, Same as \textbf{a}, but for temperature. Note that the retrieved spread would be representative of temperatures at the surface and throughout the troposphere.}
    \label{AE_Thermo}
\end{figure}

\newpage
Similar trends to the modern Earth scenario can be seen for the parameters in Figure \ref{AE_Physical} and Figure \ref{AE_Thermo}. However, the pressure constraints are notably weaker with an inability to rule out high pressure scenarios in comparison to the modern Earth results.   

\begin{figure}[H]
    \centering
    \includegraphics[scale=0.6]{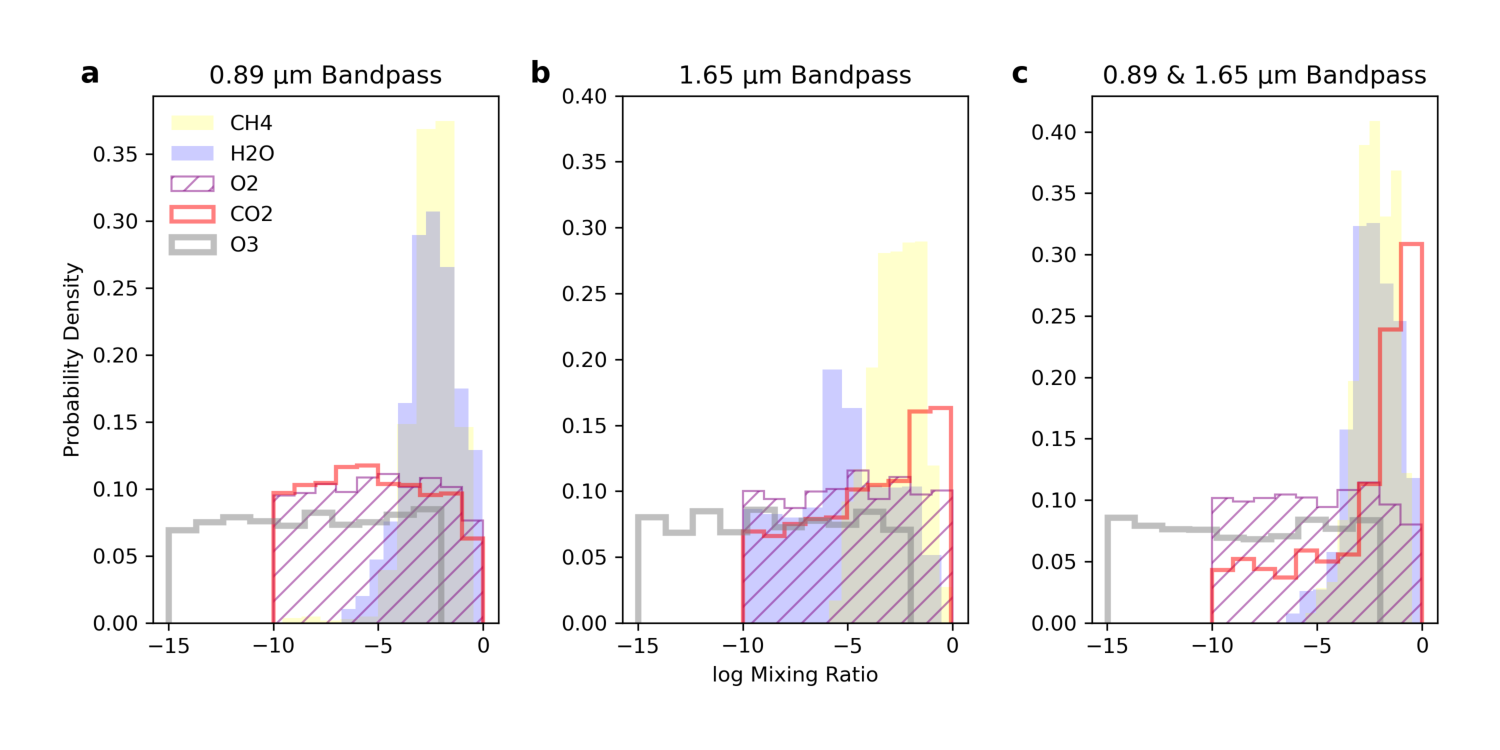}
    \caption{Biosignature Abundance Constraints Broken Down by Observational Bandpass Combinations (Archean Earth Pathway). \textbf{a}, Chemical species abundance constraints for the 0.89\,$\upmu$m bandpass. The abundance constraints for CH$_4$ (yellow filled), H$_2$O (purple filled), O$_2$ (purple hatched), CO$_2$ (red), and O$_3$ (grey) are all shown. \textbf{b}, The same as \textbf{a}, but for the singular 1.65\,$\upmu$m bandpass. \textbf{c}, The same as \textbf{a} and \textbf{b} but for the 0.89\,$\upmu$m \& 1.65\,$\upmu$m bandpass combination. Note that oxygen bearing species go unconstrained for all observational scenarios given the anoxic environment of Archean Earth.}
    \label{AE_bandpass_abund}
\end{figure}

Similar to Figure \ref{bandpass_abund} for the modern Earth decision tree pathway, Figure \ref{AE_bandpass_abund} shows species abundance constraints for each singular bandpass observation, and for the combination of the two. 

\noindent In this atmospheric scenario, all oxygen bearing species (i.e., O$_2$ and O$_3$) go unconstrained at all observational bandpass combinations due to the very low relative abundances of these species during the Archean. In the 0.89\,$\upmu$m bandpass (Figure \ref{AE_bandpass_abund}a), we can constrain both the CH$_4$ and H$_2$O in this one bandpass. It remains challenging to extract CO$_2$ abundance information even from the  1.65\,$\upmu$m singular and 0.89\,$\upmu$m \& 1.65\,$\upmu$m combined observations, which are the best case scenarios here, when at R=70. In the 0.89\,$\upmu$m \& 1.65\,$\upmu$m case the CO$_2$ abundance posterior still exhibits a statistically significant tail toward lower values making the probability to prefer low CO$_2$ abundances just as likely as CO$_2$ detections at the peak (Figure \ref{AE_bandpass_abund}c red curve). Better constraints on CO$_2$ can come from higher SNR, as demonstrated here, or higher spectral resolution. 

\section{Discussion} 
\label{sec:disc}
Overall, our decision tree retrieval analyses for both Modern and Archean Earth indicate the best constraints on each planetary parameter are achieved when maximizing the amount of spectral data we retrieve on. In other words, simulations that retrieved on a highest number of combined bandpasses (e.g., the ``All three" combination for modern Earth and the 0.89\,$\upmu$m \& 1.65\,$\upmu$m combination for Archean Earth) generally provided more contextual information that thus led to higher quality constraints on a given parameter, even when a given gas did not have absorption features in all of the bandpasses being combined. By design, the observations for single bandpasses that contained absorption features for a given species also generally provided good constraints for that species. For example, the 0.89\,$\upmu$m bandpass does well at constraining the H$_2$O abundance, because of the multiple H$_2$O features to retrieve on in this wavelength range. However, when using that same bandpass on its own to retrieve the modern Earth O$_2$ abundance, it does poorly due to a lack of O$_2$ features within that wavelength range.

Stepping through the decision tree for the modern Earth case, H$_2$O is able to be detected and even constrained to within an order of magnitude for the ``All three" combination to 1-$\sigma$ uncertainty (Table \ref{tab:ME_1sig}). We saw that for the 0.89\,$\upmu$m bandpass, including that weaker 0.82\,$\upmu$m H$_2$O feature in addition to the stronger 0.94\,$\upmu$m feature is useful for better constraining the H$_2$O abundance. This is because the depth of that weaker feature allows the retrieval model to rule out high H$_2$O abundance scenarios, which are limited by the signal strength (or depth) of that feature. Given the low abundance of CH$_4$ for modern Earth, it is not detected in the initial bandpass observation. Obtaining H$_2$O abundance constraints will inform interpretation of the actual habitability of a given potentially habitable target orbiting in the habitable zone of its host star. A non-detection of CH$_4$ in this first observation means the next observation should prioritize a search for O$_2$ as the planet is unlikely to be an Archean Earth analog. Using the ``All three" bandpass combination, gas abundance constraints for O$_2$ are (to 1-$\sigma$ confidence) within an order of magnitude of the true value. The 0.75\,$\upmu$m \& 0.89\,$\upmu$m bandpass combination also provides excellent constraints on the O$_2$ abundance, allowing us to move on to the next level of the decision tree which is conducting a search for CH$_4$ and CO$_2$ in the NIR. Both CH$_4$ and CO$_2$ remained difficult to constrain for all bandpass combinations, which was to be expected. This is due to the low relative abundances of these species for a modern Earth-like atmospheric scenario, and the correspondingly small features caused by these gases in the assumed HWO wavelength range. From the CH$_4$ posteriors, we can infer an upper limit on the abundance, and rule out an Archean Earth-like scenario with a high CH$_4$ abundance to 1-$\sigma$ level confidence for all the observational combinations (1-$\sigma$ spreads shown in Table \ref{tab:ME_1sig}). While the retrievals only ruled out very high CO$_2$ concentrations on the order of $\sim$ 10\% of the atmosphere, this can be useful in ruling out scenarios where CO$_2$-dominated atmospheres lead to the abiotic photochemical production of O$_2$ or O$_3$ \citep{Segura_2007,Domagal_Goldman_2014,Tian_2014}. The final observational priority is to constrain the atmospheric pressure. Atmospheric pressure is intentionally generalized in the decision tree without a specific wavelength range indicated. While one could tune a dedicated bandpass on the decision tree to encompass, for example, the Rayleigh scattering slope, we show here that pressure constraints can also be derived at various wavelengths from the widths of spectral features that are broadened due to pressure. The marginal posterior distributions for pressure rule out $< 0.1$ bar scenarios to 1-$\sigma$ confidence for all the observational combinations except for the 0.75\,$\upmu$m singular bandpass (1-$\sigma$ spreads shown in Table \ref{tab:ME_1sig}), helping rule out low pressure O$_2$ false positive cases. This indicates that the final categorization for the planet would be ``modern Earth" as opposed to ``Ambiguous" and an accurate characterization for our modern Earth abundance scenario.    

Going through the decision tree for the Archean Earth case, observations would again start with the top of the tree, prioritizing a search for H$_2$O and CH$_4$ in the visible. H$_2$O remains detectable from the retrieval results with the 0.89\,$\upmu$m singular bandpass, but for CH$_4$, the constraints are vastly improved in comparison to the modern Earth scenario. Since H$_2$O and CH$_4$ were successfully detected, a search for CO$_2$ and additional CH$_4$ features in the NIR would be next to provide additional geochemical context to interpret the CH$_4$ detection. In our retrievals, the CO$_2$ was on the cusp of detection at nominal SNR 10 observations with a resolving power of 70. With initial observations, this planetary scenario could be placed in the Ambiguous planet category and flagged as a potentially key target for follow-up. We performed simulated follow-up observations at an increased SNR of 20 with the singular 1.65\,$\upmu$m bandpass. Those results yielded significantly improved abundance constraints on the CO$_2$, which were able to better rule out low CO$_2$ abundance scenarios. In practice, observations at increased SNR, or increased resolving power could help strengthen the CO$_2$ inference such that this abundance scenario could be characterized as Archean Earth-like.

A notable pathway on the decision tree that will be further explored in future work is the Proterozoic Earth Pathway, which is key to contextualizing potential biosignature detections made in the UV wavelength range. In planetary scenarios where atmospheric O$_2$ is present but difficult to detect, the 0.255\,$\upmu$m bandpass (which extends from 0.23\,--\,0.28\,$\upmu$m) incorporated in the decision tree would play an important role in characterizing less oxygenated planets. In these instances, making observations in the UV would be the only avenue to detecting oxygen bearing species (e.g., O$_3$) within the proposed HWO wavelength range. As a whole, this decision tree is designed to categorize planets based on Earth through time because Earth is our only example of a habitable and inhabited world. Thus, this decision tree is designed to provide preliminary classifications based on the best-studied biosignatures expressed by Earth over its history and flag planets for detailed follow-up. However, it is important to acknowledge that the current iteration of the decision tree may not be able to identify alternative biospheres different from Earth. Future work can expand or rework the current decision tree framework to test other hypotheses, including alternative biosignatures and their false positives.

In stepping through these pathways on the decision tree, our results validate the prioritization of the 0.89\,$\upmu$m bandpass at the top of the decision tree since we can potentially constrain water and a key biosignature species in a single observation. Obtaining abundance constraints on multiple species in one observation enhances our ability to establish planetary habitability (via an H$_2$O detection) and perform more efficient preliminary exoplanet characterization studies of a given target. NIR observations with the 1.65\,$\upmu$m bandpass provide important contextual climate information through abundance inferences of key species like CO$_2$, which could help guard against misinterpretations of Archean Earth-like and CO$_2$ dominated planets \citep{Damiano_Hu_2022}. Future telescope architectures also have the potential to leverage parallel observations (in contrast to sequential observations) for increased observational efficiency. For example, with the currently assumed divisions between the visible and NIR coronagraph channels, a search for H$_2$O and/or CH$_4$ in the visible could occur simultaneously with a search for CH$_4$ and CO$_2$ in the NIR, which would be the ideal setup for classifying a potentially Archean Earth-like planet. Conversely, if the long wavelength cutoff for the visible channel were moved shortward such that the 0.75\,$\upmu$m and 0.89\,$\upmu$m bandpasses were in separate channels, then a search for H$_2$O/CH$_4$ and O$_2$ could be done in parallel, which would be more useful for a modern Earth scenario. Future trade studies on optimizing the coronagraph channel wavelength ranges will be important to consider as mission development continues. 

Another key result of this work was our ability to constrain various planetary parameters that did not have a dedicated observational bandpass in the decision tree. For both atmospheric cases, and all the simulated observational scenarios, characteristic atmospheric temperature (T$_0$) exhibited posterior distributions that were peaked near the surface temperature (288\,K for modern Earth and 296\,K for Archean Earth) with statistically significant tails extending toward high values. These constraints largely arise from water vapor opacities, which exhibit temperature sensitivity and prior work has shown similar trends \citep{Young_2023,Barrientos_2023,Robinson_2023_rfast_PSJ}. However, it is worth noting that the constraints found here are poorer than the aforementioned studies suggesting bandpasses encompassing H$_2$O features in the NIR, which are not present in this iteration of the decision tree, are necessary if the aim is to improve the atmospheric temperature constraints further. In regards to other parameters, we also saw surprisingly good constraints on the planetary radius for both atmospheric cases and all the simulated observations produced posteriors peaked at 1 R$\oplus$ with tails extending toward high values. These radius distributions all rule out (to 1-$\sigma$ confidence) planets $> 3$ R$\oplus$ (Table \ref{tab:ME_1sig}). Strong radius constraints in general can arise when the geometric albedo and/or observational phase angle are well-known. Our retrieval observations assume a fixed observational phase angle representative of what one would expect for a planet observed in quadrature/gibbous phase and is the most likely cause of the strong radius constraints. Additionally, we assume the the orbital distance of the planet is known and fixed at 1 AU which also leads to improved radius constraints. Taken together, this implies that if the scattering environment of the planet is well characterized along with the phase angle geometry and the orbital distance, that would provide sufficient information to provide constraints on planetary radius. While planetary mass was included as a retrieved parameter in these analyses, mass went unconstrained in all observational scenarios. This was to be expected given that mass (equivalent to surface gravity) has been seen to be difficult to constrain in reflected light \citep{Feng_2018}. Mass estimates could be improved with astrometry or radial velocity measurements that could be incorporated into the mass prior of the retrieval. Additional parameters that where included in the retrieval but did not have dedicated bandpasses in the decision tree were the cloud parameters (cloud thickness, cloud top pressure, cloud optical depth, and cloudiness fraction). Constraining these parameters would be a crucial aspect to exoplanet characterization because they could indicate potentially Earth-like climatic states. Cloud optical depth for example, is best constrained in the NIR with the 1.65\,$\upmu$m bandpass or the ``All three" combination (see Table \ref{tab:ME_1sig} and Table \ref{tab:AE_1sig}).  

The decision tree takes into consideration observations that may potentially rule out known false-positive scenarios. Mechanisms for generating O$_2$ abiotically, for example, are well studied in the literature \citep{Harman_2018,Domagal_Goldman_2014,Luger_Barnes_2015,Wordsworth_2014}. On the decision tree, high O$_2$ abundances lead to the dead and rocky category to represent instances where large volumes of O$_2$ are being generated from extensive photodissociation of water vapor and that then leads to atmospheric escape of the hydrogen leaving the atmospheric O$_2$ behind. This false positive mechanism generally applies to planets orbiting M dwarfs since these stars typically generate higher UV fluxes in comparison to sun-like stars. Additionally, M dwarfs go through an extended super-luminous pre-main sequence evolutionary phase that can drive off volatiles and create high levels of O$_2$ from photodissociation of H$_2$O \citep{Luger_Barnes_2015}. Another O$_2$ false-positive pathway in the decision tree is at the pressure branch. The 0.1 bar threshold for categorization represents an atmospheric regime in which a limited abundance of non-condensable species leads to a buildup of water vapor at high altitudes (due to lack of an atmospheric cold trap) and that H$_2$O is then susceptible to photolysis \citep{Wordsworth_2014}.

The decision tree retrieval analyses performed in this work are key precursor studies to the development of the upcoming Habitable Worlds Observatory and developing direct imaging observational strategies for exoplanet characterization in general. We have demonstrated the utility of the decision tree and the planetary parameter constraints that can be achieved with simulated nominal SNR 10 observations in a given 20\% bandpass at fixed resolving power. The decision tree framework has laid out a step-by-step guide for how one might prioritize observations for characterizing Earth-like planets orbiting sun-like stars but one could also expand on or adapt the decision tree based on the scientific motivations for a given observation.

\section{Conclusion} 
\label{sec:conc}
Future exoplanet characterization efforts will rely on our ability to carry out observations efficiently while also maximizing the contextual information derived from them. Coronagraph instrumentation is limited to observing fractions of a full 0.2--1.8\,$\upmu$m spectrum at a given time, but critically, we show that meaningful atmospheric inferences can be performed regardless through strategically placed spectroscopic bandpasses. The observational decision tree framework presented in this work can provide a useful roadmap for categorizing Earth-like exoplanets in a minimal number of observations. We have demonstrated the utility of the decision tree using retrieval analyses following the pathways for both the modern Earth and Archean Earth. In both atmospheric scenarios, we found that not only can we achieve constraints on biosignatures and habitability markers prioritized within the decision tree, we can also constrain additional planetary parameters not outlined in the decision tree (e.g., planetary radius, global surface pressure, and atmospheric temperature). We simulated SNR 10 observations with every singular and combination of bandpasses for each atmospheric pathway and tabulated the quality of each parameter constraint for each combination. We found that the ``All three" bandpass combination (consisting of the 0.75\,$\upmu$m bandpass from 0.68\,--\,0.81\,$\upmu$m, the 0.89\,$\upmu$m bandpass from 0.81\,--\,0.97\,$\upmu$m, and the 1.65\,$\upmu$m bandpass from 1.50\,--\,1.80\,$\upmu$m) for modern Earth and the ``0.89\,$\upmu$m \& 1.65\,$\upmu$m" bandpass combination for Archean Earth provided the best constraints for each of the retrieval parameters. Overall, we show that the decision tree can be implemented with retrieval analyses and give meaningful insights to developing observational strategies for characterizing Earth-like exoplanets.


\section*{acknowledgments}
\noindent All authors acknowledge support from internally sourced funding from the NASA GSFC Sellers Exoplanet Environments Collaboration. JC and STB acknowledge support from the Center for Research and Exploration in Space Science and Technology at NASA, Grant No.~ 80GSFC21M0002. AVY would like to acknowledge support from the from the NASA Exobiology Program, Grant No.~80NSSC18K0349 as well as the NASA Fellowship Activity Program awarded through Grant No.~80NSSC21K2064. AVY also acknowledges support from the NASA Pathways Intern Program administered by the local Pathways Office at NASA Goddard.  
\appendix


    \begin{longtable}{c c c c c c c}
    \caption{List of Parameters and 1-sigma spreads for modern Earth retrieval combinations}\\
    \hline
    \hline
    Parameter Name & Description & Input Value & Flat Prior & Obs. Bandpass & 1-sigma \\
    \hline
         log O$_2$ &  Molecular oxygen mixing ratio & -0.677 & [-10,0] & 0.75 & $-1.311_{-1.674}^{+0.908}$\\
         & & & & 0.89 & $-5.136_{-3.421}^{+3.332}$\\
         & & & & 1.65 & $-4.977_{-3.515}^{+3.336}$\\
         & & & & 0.75 \& 0.89 & $-0.609_{-1.210}^{+0.472}$\\
         & & & & 0.75 \& 1.65 & $-1.013_{-3.536}^{+0.749}$\\
         & & & & 0.89 \& 1.65 & $-5.160_{-3.354}^{+3.443}$\\
         & & & & All three & $-0.590_{-0.892}^{+0.425}$\\
         
         log H$_2$O &  Water vapor mixing ratio & -2.522 & [-10,0] & 0.75 & $-5.126_{-3.287}^{+2.262}$\\
         & & & & 0.89 & $-1.795_{-0.894}^{+1.053}$\\
         & & & & 1.65 & $-2.097_{-1.401}^{+1.177}$\\
         & & & & 0.75 \& 0.89 & $-2.438_{-0.638}^{+0.569}$\\
         & & & & 0.75 \& 1.65 & $-2.999_{-1.283}^{+1.013}$\\
         & & & & 0.89 \& 1.65 & $-2.166_{-1.151}^{+1.158}$\\
         & & & & All three & $-2.324_{-0.572}^{+0.535}$\\

         log CO$_2$ &  Carbon dioxide mixing ratio & -3.397 & [-10,0] & 0.75 & $-5.061_{-3.289}^{+3.249}$\\
         & & & & 0.89 & $-5.212_{-3.307}^{+3.244}$\\
         & & & & 1.65 & $-5.256_{-3.197}^{+2.679}$\\
         & & & & 0.75 \& 0.89 & $-5.059_{-3.426}^{+3.409}$\\
         & & & & 0.75 \& 1.65 & $-5.359_{-3.142}^{+2.968}$\\
         & & & & 0.89 \& 1.65 & $-5.093_{-3.353}^{+3.366}$\\
         & & & & All three & $-4.214_{-3.830}^{+2.187}$\\

         log O$_3$ &  Ozone mixing ratio & -6.154 & [-10,-2] & 0.75 & $-7.070_{-1.891}^{+1.372}$\\
         & & & & 0.89 & $-5.708_{-2.869}^{+2.493}$\\
         & & & & 1.65 & $-5.595_{-3.042}^{+2.749}$\\
         & & & & 0.75 \& 0.89 & $-7.291_{-1.797}^{+1.633}$\\
         & & & & 0.75 \& 1.65 & $-6.846_{-2.032}^{+1.385}$\\
         & & & & 0.89 \& 1.65 & $-6.198_{-2.656}^{+2.655}$\\
         & & & & All three & $-7.134_{-1.852}^{+1.679}$\\

         log CH$_4$ &  Methane mixing ratio & -5.698 & [-10,0] & 0.75 & $-6.156_{-2.587}^{+2.656}$\\
         & & & & 0.89 & $-6.118_{-2.759}^{+2.743}$\\
         & & & & 1.65 & $-7.237_{-1.862}^{+2.010}$\\
         & & & & 0.75 \& 0.89 & $-6.301_{-2.499}^{+2.707}$\\
         & & & & 0.75 \& 1.65 & $-7.447_{-1.785}^{+1.953}$\\
         & & & & 0.89 \& 1.65 & $-7.324_{-1.776}^{+2.080}$\\
         & & & & All three & $-7.282_{-1.887}^{+1.962}$\\

         log P$_0$ (log Pa) & Surface pressure & 5.004 & [0,8] & 0.75 & $5.658_{-0.859}^{+1.046}$\\
         & & & & 0.89 & $4.682_{-0.718}^{+0.703}$\\
         & & & & 1.65 & $6.064_{-1.094}^{+1.469}$\\
         & & & & 0.75 \& 0.89 & $5.119_{-0.550}^{+0.569}$\\
         & & & & 0.75 \& 1.65 & $5.533_{-0.784}^{+0.949}$\\
         & & & & 0.89 \& 1.65 & $5.434_{-1.008}^{+1.074}$\\
         & & & & All three & $5.399_{-0.563}^{+0.599}$\\

         T$_0$ (K) & Atmospheric temperature & 206\,--\,288 & [100,1000] & 0.75 & $376.4_{-185.8}^{+330.6}$\\
         & & & & 0.89 & $265.1_{-75.85}^{+145.1}$\\
         & & & & 1.65 & $273.4_{-104.4}^{+193.0}$\\
         & & & & 0.75 \& 0.89 & $286.6_{-83.75}^{+121.4}$\\
         & & & & 0.75 \& 1.65 & $386.1_{-216.1}^{+275.4}$\\
         & & & & 0.89 \& 1.65 & $263.4_{-70.07}^{+105.3}$\\
         & & & & All three & $278.7_{-72.91}^{+134.0}$\\

         log A$_{\rm s}$ &  Surface albedo & -1.301 & [-2,0] & 0.75 & $-0.716_{-0.672}^{+0.484}$\\
         & & & & 0.89 & $-1.060_{-0.570}^{+0.591}$\\
         & & & & 1.65 & $-1.418_{-0.433}^{+0.717}$\\
         & & & & 0.75 \& 0.89 & $-0.698_{-0.464}^{+0.495}$\\
         & & & & 0.75 \& 1.65 & $-1.089_{-0.535}^{+0.646}$\\
         & & & & 0.89 \& 1.65 & $-1.405_{-0.396}^{+0.658}$\\
         & & & & All three & $-1.072_{-0.545}^{+0.642}$\\

         R$_{\rm p}$ (R$_\oplus$) &  Planetary radius & 1.0 & [0.1,10] & 0.75 & $1.024_{-0.349}^{+0.807}$\\
         & & & & 0.89 & $1.538_{-0.705}^{+1.388}$\\
         & & & & 1.65 & $1.126_{-0.303}^{+0.587}$\\
         & & & & 0.75 \& 0.89 & $1.077_{-0.416}^{+0.620}$\\
         & & & & 0.75 \& 1.65 & $1.058_{-0.331}^{+0.772}$\\
         & & & & 0.89 \& 1.65 & $1.143_{-0.348}^{+1.229}$\\
         & & & & All three & $1.020_{-0.281}^{+0.645}$\\

         M$_{\rm p}$ (M$_\oplus$) &  Planetary mass & 1.0 & [0.1,10] & 0.75 & $5.455_{-3.271}^{+3.135}$\\
         & & & & 0.89 & $5.613_{-3.502}^{+3.108}$\\
         & & & & 1.65 & $6.026_{-3.397}^{+2.689}$\\
         & & & & 0.75 \& 0.89 & $5.072_{-3.665}^{+3.372}$\\
         & & & & 0.75 \& 1.65 & $5.600_{-3.360}^{+2.955}$\\
         & & & & 0.89 \& 1.65 & $5.696_{-3.322}^{+2.976}$\\
         & & & & All three & $5.258_{-3.338}^{+3.192}$\\

         log $\Delta {\rm P}_{\rm c}$ (log Pa) &  Cloud thickness & 4.0 & [0,8] & 0.75 & $2.717_{-1.866}^{+2.005}$\\
         & & & & 0.89 & $2.355_{-1.525}^{+1.616}$\\
         & & & & 1.65 & $3.364_{-2.325}^{+2.104}$\\
         & & & & 0.75 \& 0.89 & $2.433_{-1.812}^{+2.433}$\\
         & & & & 0.75 \& 1.65 & $2.735_{-1.904}^{+2.114}$\\
         & & & & 0.89 \& 1.65 & $2.918_{-1.945}^{+1.992}$\\
         & & & & All three & $3.083_{-2.138}^{+1.980}$\\

         log P$_{\rm t}$ (log Pa) &  Cloud top pressure & 4.778 & [0,8] & 0.75 & $2.748_{-1.845}^{+1.996}$\\
         & & & & 0.89 & $2.379_{-1.588}^{+1.576}$\\
         & & & & 1.65 & $4.630_{-1.714}^{+0.801}$\\
         & & & & 0.75 \& 0.89 & $2.613_{-1.988}^{+1.929}$\\
         & & & & 0.75 \& 1.65 & $3.088_{-2.155}^{+1.855}$\\
         & & & & 0.89 \& 1.65 & $3.890_{-2.425}^{+1.144}$\\
         & & & & All three & $3.697_{-2.560}^{+1.204}$\\

         log $\tau_{\rm c}$ &  Optical depth & 1.0 & [-3,3] & 0.75 & $0.099_{-2.151}^{+1.933}$\\
         & & & & 0.89 & $0.079_{-2.034}^{+1.892}$\\
         & & & & 1.65 & $1.723_{-0.737}^{+0.870}$\\
         & & & & 0.75 \& 0.89 & $0.0_{-1.940}^{+1.922}$\\
         & & & & 0.75 \& 1.65 & $1.167_{-0.433}^{+0.985}$\\
         & & & & 0.89 \& 1.65 & $0.988_{-0.530}^{+0.535}$\\
         & & & & All three & $1.181_{-0.316}^{+0.958}$\\

         log f$_{\rm c}$ &  Cloud fraction & -0.301 & [-3,0] & 0.75 & $-1.545_{-0.969}^{+1.051}$\\
         & & & & 0.89 & $-1.505_{-0.994}^{+1.062}$\\
         & & & & 1.65 & $-0.408_{-0.434}^{+0.285}$\\
         & & & & 0.75 \& 0.89 & $-1.427_{-1.063}^{+1.101}$\\
         & & & & 0.75 \& 1.65 & $-0.544_{-0.680}^{+0.378}$\\
         & & & & 0.89 \& 1.65 & $-0.462_{-1.001}^{+0.352}$\\
         & & & & All three & $-0.525_{-0.621}^{+0.381}$\\
    \hline
    \label{tab:ME_1sig}
    \end{longtable}

\begin{table}[H]
\caption{List of Parameters and 1-sigma spreads for Archean Earth retrieval combinations \textbf{*The CO$_2$ abundance was not well detected in the SNR 10 simulated observations and we refrain from reporting their 1-$\sigma$ intervals as they were likely subject to being influenced by the priors.}}
    \centering
    \begin{tabular}{c c c c c c c}
    \hline
    \hline
    Parameter Name & Description & Input Value & Flat Prior & Obs. Bandpass & 1-sigma \\
    \hline
         log O$_2$ &  Molecular oxygen mixing ratio & -8 & [-10,0] & 0.89 & $-5.017_{-3.319}^{+3.136}$\\
         & & & & 1.65 & $-4.839_{-3.504}^{+3.192}$\\
         & & & & 0.89 \& 1.65 & $-5.130_{-3.281}^{+3.259}$\\
         
         log H$_2$O &  Water vapor mixing ratio & -1.657 & [-10,0] & 0.89 & $-2.457_{-1.191}^{+1.301}$\\
         & & & & 1.65 & $-5.273_{-2.830}^{+2.732}$\\
         & & & & 0.89 \& 1.65 & $-2.262_{-1.092}^{+1.169}$\\
         
         \textbf{*}log CO$_2$ &  Carbon dioxide mixing ratio & -1.397 & [-10,0] & 0.89 & -\\
         & & & & 1.65 & -\\
         & & & & 1.65 (SNR 20) & $-1.211_{-1.412}^{+0.697}$\\
         & & & & 0.89 \& 1.65 & -\\

         log O$_3$ &  Ozone mixing ratio & -13.045 & [-10,-2] & 0.89 & $-8.285_{-4.517}^{+4.371}$\\
         & & & & 1.65 & $-8.558_{-4.422}^{+4.487}$\\
         & & & & 0.89 \& 1.65 & $-8.556_{-4.504}^{+4.584}$\\

         log CH$_4$ &   Methane mixing ratio & -2.522 & [-10,0] & 0.89 & $-2.329_{-0.960}^{+0.892}$\\
         & & & & 1.65 & $-2.655_{-1.315}^{+1.174}$\\
         & & & & 0.89 \& 1.65 & $-2.198_{-0.865}^{+0.955}$\\

         log P$_0$ (log Pa) &  Surface pressure & 4.993 & [0,8] & 0.89 & $5.340_{-0.942}^{+1.248}$\\
         & & & & 1.65 & $6.073_{-1.143}^{+1.176}$\\
         & & & & 0.89 \& 1.65 & $5.775_{-0.914}^{+1.501}$\\

         T$_0$ (K) & Atmospheric temperature & 220\,--\,296 & [100,1000] & 0.89 & $267.3_{-84.36}^{+149.0}$\\
         & & & & 1.65 & $340.8_{-175.7}^{+335.4}$\\
         & & & & 0.89 \& 1.65 & $304.3_{-70.20}^{+104.9}$\\

         log A$_{\rm s}$ &  Surface albedo & -1.301  & [-2,0] & 0.89 & $-1.014_{-0.620}^{+0.624}$\\
         & & & & 1.65 & $-1.494_{-0.373}^{+0.699}$\\
         & & & & 0.89 \& 1.65 & $-1.170_{-0.511}^{+0.596}$\\

         R$_{\rm p}$ (R$_\oplus$) &  Planetary radius & 1.0 & [0.1,10] & 0.89 & $1.329_{-0.579}^{+1.122}$\\
         & & & & 1.65 & $1.654_{-0.722}^{+0.857}$\\
         & & & & 0.89 \& 1.65 & $1.278_{-0.481}^{+0.800}$\\

         M$_{\rm p}$ (M$_\oplus$) &  Planetary Mass & 1.0 & [0.1,10] & 0.89 & $5.346_{-3.286}^{+3.206}$\\
         & & & & 1.65 & $5.405_{-3.537}^{+3.249}$\\
         & & & & 0.89 \& 1.65 & $5.435_{-3.369}^{+3.103}$\\

         log $\Delta$ {\rm P}$_{\rm c}$ (log Pa) &  Cloud thickness & 4.0 & [0,8] & 0.89 & $2.657_{-1.849}^{+1.967}$\\
         & & & & 1.65 & $3.103_{-2.185}^{+2.076}$\\
         & & & & 0.89 \& 1.65 & $2.935_{-1.980}^{+2.041}$\\

         log P$_{\rm t}$ (log Pa) &  Cloud top pressure & 4.778 & [0,8] & 0.89 & $2.766_{-1.894}^{+1.760}$\\
         & & & & 1.65 & $3.079_{-2.045}^{+1.641}$\\
         & & & & 0.89 \& 1.65 & $4.161_{-2.177}^{+0.981}$\\

         log $\tau_{\rm c}$ &  Optical depth & 1.0 & [-3,3] & 0.89 & $0.472_{-2.284}^{+1.732}$\\
         & & & & 1.65 & $0.183_{-1.805}^{+1.789}$\\
         & & & & 0.89 \& 1.65 & $1.563_{-2.016}^{+0.957}$\\

         log f$_{\rm c}$ &  Cloud fraction & -0.301 & [-3,0] & 0.89 & $-1.351_{-1.096}^{+0.934}$\\
         & & & & 1.65 & $-1.397_{-0.962}^{+0.896}$\\
         & & & & 0.89 \& 1.65 & $-0.811_{-1.057}^{+0.541}$\\
    \hline
    \end{tabular}
    \label{tab:AE_1sig}
\end{table}


\bibliography{references}{}
\bibliographystyle{aasjournal}



\end{document}